\documentclass[a4paper,11pt]{article}
\usepackage{pos}
\usepackage{wrapfig,enumitem}

\title{Socioeconomic Impact of the Pierre Auger Observatory}

\author*[a]{Ingomar Allekotte}
\author[b]{for the Pierre Auger Collaboration,}
\author[c]{Alejandra Calvo}

\affiliation[a]{Comisión Nacional de Energía Atómica and Instituto Balseiro, \\
Centro Atómico Bariloche, 8400 San Carlos de Bariloche, 
Argentina}

\affiliation[b]{Observatorio Pierre Auger, Av.\ San Mart{\'\i}n Norte 304, 5613 Malarg\"ue, Mendoza, Argentina\\
Full author list: 
{\rm\url{https://www.auger.org/archive/authors_icrc_2025.html}}}

\emailAdd{spokespersons@auger.org}

\affiliation[c]{Comisión Nacional de Energía Atómica, Argentina}

\abstract{
The Pierre Auger Observatory has been operating in Malargüe, Province of
Mendoza, western Argentina, for over two decades, significantly advancing our
understanding of cosmic rays. Beyond its scientific mission, the installation and operation
of the Observatory has had profound social, economic, educational and cultural impact on
the local community, the region and worldwide.
More than 90\% of the Observatory's annual operational budget is invested in the region,
benefiting sectors such as tourism, hospitality, gastronomy, and local commerce.
Additionally, the presence of international visitors and collaborations has fostered a rich
cultural exchange.
Malargüe has emerged as a destination for scientific tourism, with the Observatory as a
major attraction, having welcomed over 178,000 visitors since its inauguration and
boosting the 
development of the region.
This article explores the broad direct and indirect benefits of the Auger Observatory and
the key lessons learned from this endeavor. A new International Agreement signed in
2024 ensures the continuity of the project for at least another decade, until 2035, reaffirming its
scientific, economic, and social relevance.
}

\ConferenceLogo{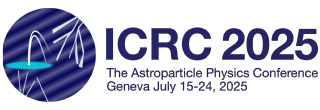}

\FullConference{39th International Cosmic Ray Conference (ICRC2025)\\
 15–24 July 2025\\
Geneva, Switzerland\\}

\begin{document}
\maketitle

\section{Introduction}
The Pierre Auger Observatory was designed and built to study cosmic rays of the highest energies \cite{Auger}. It consists of a large array with 1,660 surface detector stations (SDs), spread over an area of approximately 3,000 km$^2$, and 27 fluorescence telescopes (FDs) located at four different sites at the periphery of the surface array. This hybrid detector is supplemented with a suite of atmospheric monitoring devices. A Central Campus hosts the assembly building, the central data acquisition system, offices, storage facilities, workshops, 
shelters for vehicles, a water purification plant, and a visitor center. 

The Observatory is located on a large plain in the Departments of Malargüe and San Rafael, in the Province of Mendoza, western Argentina. The Central Campus lies on a property of 25,000 m$^2$ which belongs to the Province of Mendoza,  only 500 m from the center of the city of Malargüe. 

The Pierre Auger Observatory was conceived in 1991. In 1995, in a 6-month workshop held at Fermilab, USA, the objectives, design and bases for the construction were established, culminating with the edition of the project's first design report. 
After some years of strengthening the collaboration, and advancing the design and specifications, the groundbreaking ceremony took place in 1999, during which the ``International Agreement for the Organization, Management and Funding for the Operation of the Pierre Auger Observatory'' was signed. 

In the following years, from 2000 to 2003, the collaboration built and operated an Engineering Array \cite{EngineeringArray} with 32 water Cherenkov detector (WCD) stations and two fluorescence telescopes, to prove the concept and the feasibility of the full-scale project. Construction of the full array followed, and was completed in 2008. Regular data taking for physics analyses had already started in January 2004. 

In 2015, the international agreement was extended and a proposal to upgrade the Observatory started to take shape. The upgrade, called AugerPrime \cite{AugerPrime}, consists in enhancing the WCD array with scintillator surface detectors (SSDs) and radio detectors (RDs), adding a small PMT to increase the dynamic range of the WCDs, and renewing the electronics for all stations. 
Additionally, an array of Underground Muon Detectors (UMDs), deployed in a smaller area (23 km$^2$) of the array, complete the upgrade. The installation of the SSD, RD and new electronics was finalized in November 2024, just in time for the signature of a further extension of the international agreement which guarantees continuity of the project until December 31st, 2035. 

Presently, the Pierre Auger Observatory is run by an international collaboration of nearly 400 members from 87 institutions and 17 countries (Argentina, Australia, Belgium, Brazil, Colombia, Czech Republic, France, Germany, Italy, Mexico, Netherlands, Poland, Portugal, Romania, Spain, Slovenia, USA\@). A total of 45 employees work at the Observatory's premises in Malargüe, including 14 professionals (physicists, engineers, etc.\@) and 22 technical staff. The administration of the Observatory is delegated to an ad hoc created entity, called Fundación Ahuekna, whose founding members are the Atomic Energy Commission of Argentina (CNEA), the Province of Mendoza and the Municipality of Malargüe.

\subsection{The local environment}

A huge scientific endeavor such as the construction and operation of the Pierre Auger Observatory has an impact on individuals, groups, and organizations at very different levels, all of which have to be addressed and their interests and concerns taken into account. The main stakeholders of the Auger Observatory are: the international and Argentinian scientific community; employees and scientists, engineers, and technicians working for the Observatory, students; 
financing Agencies; participating institutions;
Municipality of Malargüe, Province of Mendoza; local population and visitors; landowners and residents in the area of the Observatory; providers of services and goods; population in general; and regulatory agencies.

The city of Malargüe is quite isolated, being the only large city of the homonymous department (41,000 km$^2$), with a population of 33,100 inhabitants. However, the city is expanding and developing rapidly, with an increase in population of 20\% in the last 10 years. The next nearby city is San Rafael, 180 km away and with more than 210,000 inhabitants. 

Since the beginning of the Auger Project, attention was paid to its potential impact, reaching far beyond the direct benefits derived from its specific research program in basic science. Engagement with the local community was key for the success of the project. Close ties were established with the local society in Malargüe and a broad outreach program was planned from the beginning. General talks about the Observatory and its mission were offered to the general public. The Central Campus was equipped with a visitor center even before data collection began. Although in the early stages of the project it was foreseen to build a guest house for visiting scientists, finally it was decided not to do so, to promote the use of the lodging offers of the city. During construction, to raise public awareness of the project, the Auger Collaboration organized a competition among local schoolchildren to select names for the individual detectors.  

The Auger Collaboration also actively participated with a seat in the Strategic Planning Commission of the city of Malargüe for many years. 
Prompted by the Auger Observatory, the City Council of Malargüe issued an ordinance for protection of the night skies and avoidance of luminous contamination, which as a by-product leads to energy efficiency. 

Every year in November, Malargüe celebrates its date of foundation with a huge parade, in which the regular participation of the Auger scientists and employees has become a tradition,
showing the commitment of the collaboration with the happenings in the local community and providing a cultural experience for foreign collaborators. 

Technicians from the Auger Observatory, together with scientists, also make regular visits to local rural schools, providing social assistance, helping to upgrade the infrastructure, and making repairs and improvements. 

\subsection{Landowners}

The area where the Auger Project is active is owned by 101 different landowners. Agreements were signed for the installation of the FD buildings and of the surface detectors, and for the transit through the properties. Over time, many of the agreements, which were initially made for 20 years, had to be renewed. Also, with successions and sales of the properties, new agreements had to be made and negotiations with the new landowners were due. The Auger Project pays a yearly fee of approximately 20 US dollars per detector, and an amount of 400 USD dollars per year for each of the four FD sites. 
Additionally, the Auger Collaboration has acquired unrestricted access to a property of 30 km$^2$, which is used as a prototyping and test area and to host an ``infill array'', with SDs located at smaller spacing. 

\section{Scientific production and training of human resources } 

As expected for a scientific endeavor of this kind, the main evaluation parameters to assess its success, are the level at which the scientific objectives are met, the scientific production and its impact. These can be measured with the usual metrics of number of papers and their citations. Between 2004, year of the collaboration's first paper, and 2025, the Auger Collaboration has published 141 full-author-list papers in refereed journals (11 more are submitted), and hundreds of conference proceeding papers. However, it is worth emphasizing that the scientific impact of a large-scale project often transcends its original scope and objectives. In the case of the Auger Observatory, this can be assessed by observing the scientific output in areas that were not within the objectives of the project at the time of its approval, such as solar activity studies \cite{Scalers}, atmospheric studies \cite{Elves, TGFs, Aerosols}, and discoveries with a remotely operated telescope \cite{FRAM}. 

\begin{figure}
    \centering
    
\includegraphics[width=0.8\linewidth]{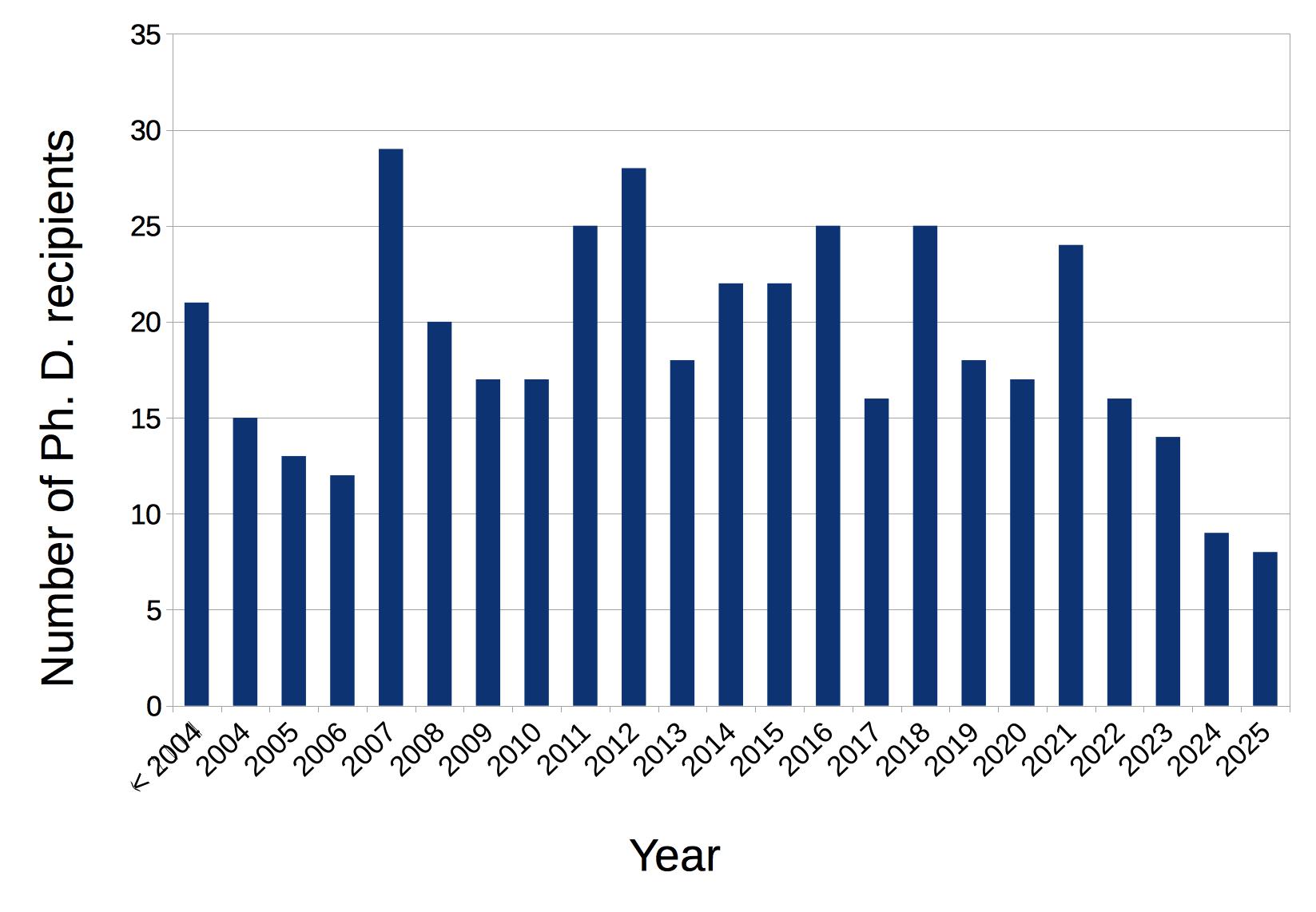}
    \caption{Completed doctoral theses in the Auger Collaboration, per year.}
    \label{fig:phds}

\end{figure}

Training of human capital is also of utmost relevance. A total of 419 students have completed their doctoral thesis working in the Auger Collaboration, and  88 Ph.~D.\ students are presently pursuing their degree (see Fig.~\ref{fig:phds}). A double degree doctoral program was established between Argentina (Universidad Nacional de San Martín) and Germany (Karlsruhe Institute of Technology), of which 27 students have already graduated and 11 theses are in progress.

\section{Infrastructure for other scientific projects}

The existence of a facility like the Pierre Auger Observatory in a remote area, with the possibility of providing infrastructure, trained technicians, and logistics support, has facilitated the installation and operation of different instruments, and provided support for other projects. With support from the Auger Observatory, different prototypes were installed and tested, such as the AERA radio-array \cite{AERA}, and projects for detection of air showers in the GHz domain, namely MIDAS, EASIER and AMBER \cite{GHZ}. FAST \cite{FAST}, a prototype array of single pixel fluorescence detectors, is benefiting from the existing infrastructure. The Auger Observatory also provides support in the installation and operation of prototype detectors for the GRAND \cite{GRAND} and IceCube \cite{IceCube} projects. A special mention deserves a spin-off of the Auger Project, the LAGO Project \cite{LAGO}, a Latin-American collaboration to install water-Cherenkov detectors at high altitudes at different latitudes, for the study of background radiation , but with a very strong emphasis on training students and young researchers in the construction and operation of radiation detectors. 

Support was also given to other projects in the area, such as for the maintenance of reflector cones for the calibration of satellite cameras from the Argentinian Space Agency CONAE, support for the maintenance of instruments for aerosol studies, and for seismic and geological research (see, for example \cite{seismic}). More recently, the detection of lasers from satellites has prompted collaborations with the space missions Aeolus and EarthCARE \cite{Satellites}.

\section{Economic Impact}

Construction of the Pierre Auger Observatory required an investment of 53 million US dollars (MUSD). For the AugerPrime upgrade the budget was of an additional 15 MUSD. 
In the construction phase, costs included the erection of the Central Campus (with assembly and office buildings) and four FD telescope buildings, providing work opportunities to local companies. These also participated in deployment of detectors, ground preparation, and provision of components and services with high quality requirements. For example, all of the rotomolded detector enclosures were provided by Mexico, Brazil, and Argentina, enhancing the capability of Latin-American rotomolding companies to produce high-quality products. For the fabrication of liners (water enclosures for the WCDs), a factory was established at the National Technological University (Mendoza branch), providing not only job opportunities but also training of students and young professionals in setting up a factory and implementing a sophisticated quality assurance program. 

The yearly budget for Operations and Maintenance of the Observatory is approximately 2 MUSD. This amount is supplemented by institutions of the Auger Collaboration with salaries and travel support for scientific collaborators, support for institutional activities on site, funds for sending shifters for FD data taking, and additional contributions (like spares or replacement equipment provided by some countries). Argentina, being the host country, provides its contribution in kind, mainly through direct hiring of staff by CNEA and CONICET. More than 90\% of the yearly operational budget is spent locally in the city of Malargüe, distributed as follows: 60\% for salaries of local staff, 15\% for maintenance (mainly vehicles), 15\% for services (city water, gas, electricity, internet and telephone services) and 10\% for administration, insurances and taxes. 

Typically, two collaboration meetings are held every year in the city of Malargüe. The collaboration has benefited from the existence of a modern Convention Center owned by the city, which was built at the time of construction of the Observatory. 
Each Auger Collaboration meeting is attended by more than 100 collaborators, who normally stay in town for one week or longer. Also, throughout the year, visiting scientists spend time at the Observatory, as well as shifters and other occasional visitors. These visits have a noticeable impact on the local economy (hotels, guest houses, restaurants, bars and touristic service providers), especially because the higher affluence of Auger collaborators occurs outside the normal touristic seasons. 

Infrastructure built for the Observatory also contributed to local economic development. A power line which was constructed for the Auger Observatory is used by local landowners and inhabitants along the trace of the line. A project to provide internet to the Observatory was funded by the European Union with 900,000 EUR and received an additional local matching finance from the local government. Wi-fi is provided directly by the Observatory to some land inhabitants in remote regions. 

One local school in Malargüe, ``Escuela James W. Cronin'', named after the founder and inspirer of the Auger Observatory, received a donation of 500,000 USD from a charitable funder for the renewal and extension of the school's building. Collaborators have also contributed with donations to improve the equipment and provide didactic material to different local schools.

\section{Social and Cultural Impact}

\begin{figure}
    \centering
    \includegraphics[width=0.6\linewidth]{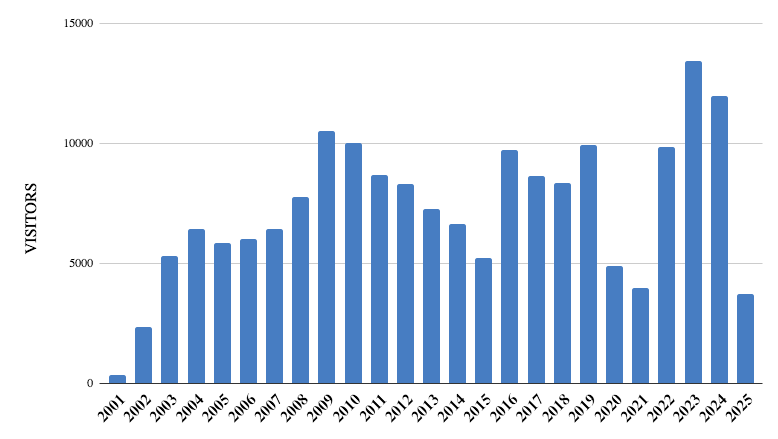}
    \caption{Visitors at the Auger Visitor Center in Malargüe (including schools) per year.}
    \label{fig:visitors}

\end{figure}

The Auger Visitor Center is open to the general public 7 days a week and receives local citizens, tourists and school groups. In the last years, it has received more than 10,000 visitors per year (Fig. \ref{fig:visitors}), a large number if compared to the local population. These visits not only provide the general public with information about the activities of the Observatory but, more importantly, motivate mainly children and teenagers to acquire interest in science and technology. 

The Auger Collaboration organizes biennially a science fair for primary and secondary schoolchildren with their teachers. For these fairs, school science projects are proposed by students and teachers from Malargüe, the Province of Mendoza and even other provinces. Of these, 25 projects are selected for presentation. The fair has already had nine editions, with a large participation of students, teachers, and the general public, totaling more than 500 participants at each event. 
These science fairs also provide a good opportunity for the international collaborators, who act as judges and evaluators of the projects, to interact with the local students. The Auger Collaboration has also launched a program of school visits, dubbed ``Auger goes to School'', to enrich the knowledge about the Observatory in the city and its surroundings. 

A fellowship program with local schools allows students to participate in work practices at the Observatory, providing them with experience in a real work environment. Up to now, 92 students have participated in this program. 

Some years after the creation of the Observatory, the National University of Cuyo, together with the local authorities of Malargüe, decided to open a branch of the university in Malargüe, providing for the first time an academic offer at university level. The professionals working at the Auger Observatory were hired as professors and teaching assistants, and without their involvement, this project would not have been possible.

A fellowship program by Michigan Technological University (MTU) provides full fellowships for local students of Malargüe to pursue their studies at MTU in USA, admitting one new student every other year. Up to now, nine students from Malargüe have enrolled in this program, and five have completed their studies and are following a successful professional career. 

The installation of the Observatory has had a profound impact on the local community regarding the perception of the relevance of foreign languages. Since the commencement of activities of the Auger Collaboration in the region, the assistance to local language schools has largely increased. Language schools are often visited by international collaborators, providing practical experience in communicating with foreigners. These encounters also contribute to a rich cultural exchange. 

The installation of the Auger Observatory in Malargüe prompted the city administration to create a Planetarium, with a very modern digital projection system and with its own production of contents and international agreements for the exchange of shows. It features a projection dome and seats for 65 people. It also has a video room, exhibits, and sundial displays. The Planetarium was visited by 300,000 visitors since its creation 17 years ago.

Due to the success of the Auger Project, the European Space Agency (ESA) contacted the local Auger Collaboration in 2008 to find a site for the installation of a huge 35 m diameter antenna for communications with interplanetary missions, the Deep Space Antenna 3  \cite{DS3}. With support from the Auger Collaboration (knowledge of the region, survey of the area, meteorological data), a site nearby the Observatory was selected. Its installation provided further work opportunities for employees and contractors. 

The DSA-3 antenna, the Auger Observatory with its visitor center and the Planetarium, together with other attractions (a dinosaur park with real dinosaur footstep traces, a geologically active surrounding with volcanoes and caves, natural reserves of wildlife and birds, an
International Center for Earth Sciences) induced Malargüe to promote itself as a destination for ``science tourism'', attracting a large number of national and international visitors. 

Minor services to local population (pure water provision to hospital and technical schools, assistance with cranes and machinery in emergencies, logistic support) are a further contribution to the city and rural areas of influence of the Auger Project. Unused or obsolete material of the Observatory is being donated to local community centers and schools (for example, packaging material is being used by local day care centers to build mattresses for the children or by local population for insulation and protection of rural houses).
More recently, the Auger Observatory has been providing disused solar panels and batteries for local social purposes.

Also, due to the experience with the Auger Project, the international collaboration of the QUBIC (``Q\& U Bolometric Interferometer for Cosmology'') project decided to install its telescope in the Province of Salta, Argentina. It is envisioned that QUBIC will have also a significant social impact, following the tracks of the Auger Observatory.

International actions, like the Open Data program \cite{OpenData}, participation and organization of Masterclasses and  other academic and outreach activities (see \cite{Outreach}) augment the visibility of Malargüe and the Province of Mendoza beyond the national borders. 

\section{Conclusions}

Large scientific projects have a huge impact, particularly if they are being carried out in remote areas or small cities. Direct economic and scientific impact is easy to quantify, however a huge number of actions have an intangible impact in society and local culture and affect perception of science and scientific activities. 

The installation and operation of the Pierre Auger Observatory in Malargüe has given the city, the Province and the region a huge 
visibility, both for the science results that are being reported word-wide, but also because of the appearance in national news and outreach initiatives. In particular, large scientific projects provide an important motivation for the younger population to engage in science and technology, a much needed trend for regional development.

\clearpage

\section*{The Pierre Auger Collaboration}
\small

\begin{wrapfigure}[8]{l}{0.11\linewidth}
\vspace{-5mm}
\includegraphics[width=0.98\linewidth]{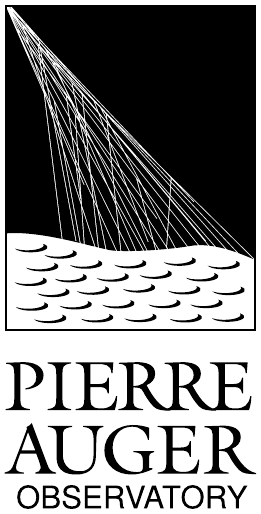}
\end{wrapfigure}
\begin{sloppypar}\noindent
A.~Abdul Halim$^{13}$,
P.~Abreu$^{70}$,
M.~Aglietta$^{53,51}$,
I.~Allekotte$^{1}$,
K.~Almeida Cheminant$^{78,77}$,
A.~Almela$^{7,12}$,
R.~Aloisio$^{44,45}$,
J.~Alvarez-Mu\~niz$^{76}$,
A.~Ambrosone$^{44}$,
J.~Ammerman Yebra$^{76}$
% ...

\end{sloppypar}

\begin{center}
\rule{0.1\columnwidth}{0.5pt}
\raisebox{-0.4ex}{\scriptsize$\bullet$}
\rule{0.1\columnwidth}{0.5pt}
\end{center}

\vspace{-1ex}
\footnotesize
\begin{description}[labelsep=0.2em,align=right,labelwidth=0.7em,labelindent=0em,leftmargin=2em,noitemsep,before={\renewcommand\makelabel[1]{##1 }}]
\item[$^{1}$] Centro At\'omico Bariloche and Instituto Balseiro (CNEA-UNCuyo-CONICET), San Carlos de Bariloche, Argentina
\item[$^{2}$] Departamento de F\'\i{}sica and Departamento de Ciencias de la Atm\'osfera y los Oc\'eanos, FCEyN, Universidad de Buenos Aires and CONICET, Buenos Aires, Argentina
\item[$^{3}$] IFLP, Universidad Nacional de La Plata and CONICET, La Plata, Argentina
\item[$^{4}$] Instituto de Astronom\'\i{}a y F\'\i{}sica del Espacio (IAFE, CONICET-UBA), Buenos Aires, Argentina
\item[$^{5}$] Instituto de F\'\i{}sica de Rosario (IFIR) -- CONICET/U.N.R.\ and Facultad de Ciencias Bioqu\'\i{}micas y Farmac\'euticas U.N.R., Rosario, Argentina
\item[$^{6}$] Instituto de Tecnolog\'\i{}as en Detecci\'on y Astropart\'\i{}culas (CNEA, CONICET, UNSAM), and Universidad Tecnol\'ogica Nacional -- Facultad Regional Mendoza (CONICET/CNEA), Mendoza, Argentina
\item[$^{7}$] Instituto de Tecnolog\'\i{}as en Detecci\'on y Astropart\'\i{}culas (CNEA, CONICET, UNSAM), Buenos Aires, Argentina
\item[$^{8}$] International Center of Advanced Studies and Instituto de Ciencias F\'\i{}sicas, ECyT-UNSAM and CONICET, Campus Miguelete -- San Mart\'\i{}n, Buenos Aires, Argentina
\item[$^{9}$] Laboratorio Atm\'osfera -- Departamento de Investigaciones en L\'aseres y sus Aplicaciones -- UNIDEF (CITEDEF-CONICET), Argentina
\item[$^{10}$] Observatorio Pierre Auger, Malarg\"ue, Argentina
\item[$^{11}$] Observatorio Pierre Auger and Comisi\'on Nacional de Energ\'\i{}a At\'omica, Malarg\"ue, Argentina
\item[$^{12}$] Universidad Tecnol\'ogica Nacional -- Facultad Regional Buenos Aires, Buenos Aires, Argentina
\item[$^{13}$] University of Adelaide, Adelaide, S.A., Australia
\item[$^{14}$] Universit\'e Libre de Bruxelles (ULB), Brussels, Belgium
\item[$^{15}$] Vrije Universiteit Brussels, Brussels, Belgium
\item[$^{16}$] Centro Brasileiro de Pesquisas Fisicas, Rio de Janeiro, RJ, Brazil
\item[$^{17}$] Centro Federal de Educa\c{c}\~ao Tecnol\'ogica Celso Suckow da Fonseca, Petropolis, Brazil
\item[$^{18}$] Instituto Federal de Educa\c{c}\~ao, Ci\^encia e Tecnologia do Rio de Janeiro (IFRJ), Brazil
\item[$^{19}$] Universidade de S\~ao Paulo, Escola de Engenharia de Lorena, Lorena, SP, Brazil
\item[$^{20}$] Universidade de S\~ao Paulo, Instituto de F\'\i{}sica de S\~ao Carlos, S\~ao Carlos, SP, Brazil
\item[$^{21}$] Universidade de S\~ao Paulo, Instituto de F\'\i{}sica, S\~ao Paulo, SP, Brazil
\item[$^{22}$] Universidade Estadual de Campinas (UNICAMP), IFGW, Campinas, SP, Brazil
\item[$^{23}$] Universidade Estadual de Feira de Santana, Feira de Santana, Brazil
\item[$^{24}$] Universidade Federal de Campina Grande, Centro de Ciencias e Tecnologia, Campina Grande, Brazil
\item[$^{25}$] Universidade Federal do ABC, Santo Andr\'e, SP, Brazil
\item[$^{26}$] Universidade Federal do Paran\'a, Setor Palotina, Palotina, Brazil
\item[$^{27}$] Universidade Federal do Rio de Janeiro, Instituto de F\'\i{}sica, Rio de Janeiro, RJ, Brazil
\item[$^{28}$] Universidad de Medell\'\i{}n, Medell\'\i{}n, Colombia
\item[$^{29}$] Universidad Industrial de Santander, Bucaramanga, Colombia
\item[$^{30}$] Charles University, Faculty of Mathematics and Physics, Institute of Particle and Nuclear Physics, Prague, Czech Republic
\item[$^{31}$] Institute of Physics of the Czech Academy of Sciences, Prague, Czech Republic
\item[$^{32}$] Palacky University, Olomouc, Czech Republic
\item[$^{33}$] CNRS/IN2P3, IJCLab, Universit\'e Paris-Saclay, Orsay, France
\item[$^{34}$] Laboratoire de Physique Nucl\'eaire et de Hautes Energies (LPNHE), Sorbonne Universit\'e, Universit\'e de Paris, CNRS-IN2P3, Paris, France
\item[$^{35}$] Univ.\ Grenoble Alpes, CNRS, Grenoble Institute of Engineering Univ.\ Grenoble Alpes, LPSC-IN2P3, 38000 Grenoble, France
\item[$^{36}$] Universit\'e Paris-Saclay, CNRS/IN2P3, IJCLab, Orsay, France
\item[$^{37}$] Bergische Universit\"at Wuppertal, Department of Physics, Wuppertal, Germany
\item[$^{38}$] Karlsruhe Institute of Technology (KIT), Institute for Experimental Particle Physics, Karlsruhe, Germany
\item[$^{39}$] Karlsruhe Institute of Technology (KIT), Institut f\"ur Prozessdatenverarbeitung und Elektronik, Karlsruhe, Germany
\item[$^{40}$] Karlsruhe Institute of Technology (KIT), Institute for Astroparticle Physics, Karlsruhe, Germany
\item[$^{41}$] RWTH Aachen University, III.\ Physikalisches Institut A, Aachen, Germany
\item[$^{42}$] Universit\"at Hamburg, II.\ Institut f\"ur Theoretische Physik, Hamburg, Germany
\item[$^{43}$] Universit\"at Siegen, Department Physik -- Experimentelle Teilchenphysik, Siegen, Germany
\item[$^{44}$] Gran Sasso Science Institute, L'Aquila, Italy
\item[$^{45}$] INFN Laboratori Nazionali del Gran Sasso, Assergi (L'Aquila), Italy
\item[$^{46}$] INFN, Sezione di Catania, Catania, Italy
\item[$^{47}$] INFN, Sezione di Lecce, Lecce, Italy
\item[$^{48}$] INFN, Sezione di Milano, Milano, Italy
\item[$^{49}$] INFN, Sezione di Napoli, Napoli, Italy
\item[$^{50}$] INFN, Sezione di Roma ``Tor Vergata'', Roma, Italy
\item[$^{51}$] INFN, Sezione di Torino, Torino, Italy
\item[$^{52}$] Istituto di Astrofisica Spaziale e Fisica Cosmica di Palermo (INAF), Palermo, Italy
\item[$^{53}$] Osservatorio Astrofisico di Torino (INAF), Torino, Italy
\item[$^{54}$] Politecnico di Milano, Dipartimento di Scienze e Tecnologie Aerospaziali , Milano, Italy
\item[$^{55}$] Universit\`a del Salento, Dipartimento di Matematica e Fisica ``E.\ De Giorgi'', Lecce, Italy
\item[$^{56}$] Universit\`a dell'Aquila, Dipartimento di Scienze Fisiche e Chimiche, L'Aquila, Italy
\item[$^{57}$] Universit\`a di Catania, Dipartimento di Fisica e Astronomia ``Ettore Majorana``, Catania, Italy
\item[$^{58}$] Universit\`a di Milano, Dipartimento di Fisica, Milano, Italy
\item[$^{59}$] Universit\`a di Napoli ``Federico II'', Dipartimento di Fisica ``Ettore Pancini'', Napoli, Italy
\item[$^{60}$] Universit\`a di Palermo, Dipartimento di Fisica e Chimica ''E.\ Segr\`e'', Palermo, Italy
\item[$^{61}$] Universit\`a di Roma ``Tor Vergata'', Dipartimento di Fisica, Roma, Italy
\item[$^{62}$] Universit\`a Torino, Dipartimento di Fisica, Torino, Italy
\item[$^{63}$] Benem\'erita Universidad Aut\'onoma de Puebla, Puebla, M\'exico
\item[$^{64}$] Unidad Profesional Interdisciplinaria en Ingenier\'\i{}a y Tecnolog\'\i{}as Avanzadas del Instituto Polit\'ecnico Nacional (UPIITA-IPN), M\'exico, D.F., M\'exico
\item[$^{65}$] Universidad Aut\'onoma de Chiapas, Tuxtla Guti\'errez, Chiapas, M\'exico
\item[$^{66}$] Universidad Michoacana de San Nicol\'as de Hidalgo, Morelia, Michoac\'an, M\'exico
\item[$^{67}$] Universidad Nacional Aut\'onoma de M\'exico, M\'exico, D.F., M\'exico
\item[$^{68}$] Institute of Nuclear Physics PAN, Krakow, Poland
\item[$^{69}$] University of \L{}\'od\'z, Faculty of High-Energy Astrophysics,\L{}\'od\'z, Poland
\item[$^{70}$] Laborat\'orio de Instrumenta\c{c}\~ao e F\'\i{}sica Experimental de Part\'\i{}culas -- LIP and Instituto Superior T\'ecnico -- IST, Universidade de Lisboa -- UL, Lisboa, Portugal
\item[$^{71}$] ``Horia Hulubei'' National Institute for Physics and Nuclear Engineering, Bucharest-Magurele, Romania
\item[$^{72}$] Institute of Space Science, Bucharest-Magurele, Romania
\item[$^{73}$] Center for Astrophysics and Cosmology (CAC), University of Nova Gorica, Nova Gorica, Slovenia
\item[$^{74}$] Experimental Particle Physics Department, J.\ Stefan Institute, Ljubljana, Slovenia
\item[$^{75}$] Universidad de Granada and C.A.F.P.E., Granada, Spain
\item[$^{76}$] Instituto Galego de F\'\i{}sica de Altas Enerx\'\i{}as (IGFAE), Universidade de Santiago de Compostela, Santiago de Compostela, Spain
\item[$^{77}$] IMAPP, Radboud University Nijmegen, Nijmegen, The Netherlands
\item[$^{78}$] Nationaal Instituut voor Kernfysica en Hoge Energie Fysica (NIKHEF), Science Park, Amsterdam, The Netherlands
\item[$^{79}$] Stichting Astronomisch Onderzoek in Nederland (ASTRON), Dwingeloo, The Netherlands
\item[$^{80}$] Universiteit van Amsterdam, Faculty of Science, Amsterdam, The Netherlands
\item[$^{81}$] Case Western Reserve University, Cleveland, OH, USA
\item[$^{82}$] Colorado School of Mines, Golden, CO, USA
\item[$^{83}$] Department of Physics and Astronomy, Lehman College, City University of New York, Bronx, NY, USA
\item[$^{84}$] Michigan Technological University, Houghton, MI, USA
\item[$^{85}$] New York University, New York, NY, USA
\item[$^{86}$] University of Chicago, Enrico Fermi Institute, Chicago, IL, USA
\item[$^{87}$] University of Delaware, Department of Physics and Astronomy, Bartol Research Institute, Newark, DE, USA
\item[] -----
\item[$^{a}$] Max-Planck-Institut f\"ur Radioastronomie, Bonn, Germany
\item[$^{b}$] also at Kapteyn Institute, University of Groningen, Groningen, The Netherlands
\item[$^{c}$] School of Physics and Astronomy, University of Leeds, Leeds, United Kingdom
\item[$^{d}$] Fermi National Accelerator Laboratory, Fermilab, Batavia, IL, USA
\item[$^{e}$] Pennsylvania State University, University Park, PA, USA
\item[$^{f}$] Colorado State University, Fort Collins, CO, USA
\item[$^{g}$] Louisiana State University, Baton Rouge, LA, USA
\item[$^{h}$] now at Graduate School of Science, Osaka Metropolitan University, Osaka, Japan
\item[$^{i}$] Institut universitaire de France (IUF), France
\item[$^{j}$] now at Technische Universit\"at Dortmund and Ruhr-Universit\"at Bochum, Dortmund and Bochum, Germany
\end{description}

\vspace{-1ex}
\footnotesize
\section*{Acknowledgments}

\begin{sloppypar}
The successful installation, commissioning, and operation of the Pierre
Auger Observatory would not have been possible without the strong
commitment and effort from the technical and administrative staff in
Malarg\"ue. We are very grateful to the following agencies and
organizations for financial support:
\end{sloppypar}

\begin{sloppypar}
Argentina -- Comisi\'on Nacional de Energ\'\i{}a At\'omica; Agencia Nacional de
Promoci\'on Cient\'\i{}fica y Tecnol\'ogica (ANPCyT); Consejo Nacional de
Investigaciones Cient\'\i{}ficas y T\'ecnicas (CONICET); Gobierno de la
Provincia de Mendoza; Municipalidad de Malarg\"ue; NDM Holdings and Valle
Las Le\~nas; in gratitude for their continuing cooperation over land
access; Australia -- the Australian Research Council; Belgium -- Fonds
de la Recherche Scientifique (FNRS); Research Foundation Flanders (FWO),
Marie Curie Action of the European Union Grant No.~101107047; Brazil --
Conselho Nacional de Desenvolvimento Cient\'\i{}fico e Tecnol\'ogico (CNPq);
Financiadora de Estudos e Projetos (FINEP); Funda\c{c}\~ao de Amparo \`a
Pesquisa do Estado de Rio de Janeiro (FAPERJ); S\~ao Paulo Research
Foundation (FAPESP) Grants No.~2019/10151-2, No.~2010/07359-6 and
No.~1999/05404-3; Minist\'erio da Ci\^encia, Tecnologia, Inova\c{c}\~oes e
Comunica\c{c}\~oes (MCTIC); Czech Republic -- GACR 24-13049S, CAS LQ100102401,
MEYS LM2023032, CZ.02.1.01/0.0/0.0/16{\textunderscore}013/0001402,
CZ.02.1.01/0.0/0.0/18{\textunderscore}046/0016010 and
CZ.02.1.01/0.0/0.0/17{\textunderscore}049/0008422 and CZ.02.01.01/00/22{\textunderscore}008/0004632;
France -- Centre de Calcul IN2P3/CNRS; Centre National de la Recherche
Scientifique (CNRS); Conseil R\'egional Ile-de-France; D\'epartement
Physique Nucl\'eaire et Corpusculaire (PNC-IN2P3/CNRS); D\'epartement
Sciences de l'Univers (SDU-INSU/CNRS); Institut Lagrange de Paris (ILP)
Grant No.~LABEX ANR-10-LABX-63 within the Investissements d'Avenir
Programme Grant No.~ANR-11-IDEX-0004-02; Germany -- Bundesministerium
f\"ur Bildung und Forschung (BMBF); Deutsche Forschungsgemeinschaft (DFG);
Finanzministerium Baden-W\"urttemberg; Helmholtz Alliance for
Astroparticle Physics (HAP); Helmholtz-Gemeinschaft Deutscher
Forschungszentren (HGF); Ministerium f\"ur Kultur und Wissenschaft des
Landes Nordrhein-Westfalen; Ministerium f\"ur Wissenschaft, Forschung und
Kunst des Landes Baden-W\"urttemberg; Italy -- Istituto Nazionale di
Fisica Nucleare (INFN); Istituto Nazionale di Astrofisica (INAF);
Ministero dell'Universit\`a e della Ricerca (MUR); CETEMPS Center of
Excellence; Ministero degli Affari Esteri (MAE), ICSC Centro Nazionale
di Ricerca in High Performance Computing, Big Data and Quantum
Computing, funded by European Union NextGenerationEU, reference code
CN{\textunderscore}00000013; M\'exico -- Consejo Nacional de Ciencia y Tecnolog\'\i{}a
(CONACYT) No.~167733; Universidad Nacional Aut\'onoma de M\'exico (UNAM);
PAPIIT DGAPA-UNAM; The Netherlands -- Ministry of Education, Culture and
Science; Netherlands Organisation for Scientific Research (NWO); Dutch
national e-infrastructure with the support of SURF Cooperative; Poland
-- Ministry of Education and Science, grants No.~DIR/WK/2018/11 and
2022/WK/12; National Science Centre, grants No.~2016/22/M/ST9/00198,
2016/23/B/ST9/01635, 2020/39/B/ST9/01398, and 2022/45/B/ST9/02163;
Portugal -- Portuguese national funds and FEDER funds within Programa
Operacional Factores de Competitividade through Funda\c{c}\~ao para a Ci\^encia
e a Tecnologia (COMPETE); Romania -- Ministry of Research, Innovation
and Digitization, CNCS-UEFISCDI, contract no.~30N/2023 under Romanian
National Core Program LAPLAS VII, grant no.~PN 23 21 01 02 and project
number PN-III-P1-1.1-TE-2021-0924/TE57/2022, within PNCDI III; Slovenia
-- Slovenian Research Agency, grants P1-0031, P1-0385, I0-0033, N1-0111;
Spain -- Ministerio de Ciencia e Innovaci\'on/Agencia Estatal de
Investigaci\'on (PID2019-105544GB-I00, PID2022-140510NB-I00 and
RYC2019-027017-I), Xunta de Galicia (CIGUS Network of Research Centers,
Consolidaci\'on 2021 GRC GI-2033, ED431C-2021/22 and ED431F-2022/15),
Junta de Andaluc\'\i{}a (SOMM17/6104/UGR and P18-FR-4314), and the European
Union (Marie Sklodowska-Curie 101065027 and ERDF); USA -- Department of
Energy, Contracts No.~DE-AC02-07CH11359, No.~DE-FR02-04ER41300,
No.~DE-FG02-99ER41107 and No.~DE-SC0011689; National Science Foundation,
Grant No.~0450696, and NSF-2013199; The Grainger Foundation; Marie
Curie-IRSES/EPLANET; European Particle Physics Latin American Network;
and UNESCO.
\end{sloppypar}


\small

\begin{thebibliography}{99}
\raggedright
\sloppy
\setlength{\itemsep}{0pt}
\small{
\bibitem{Auger} The Pierre Auger Collaboration, 
Nucl. Instrum. Meth. A 798 (2015) 172-213

\bibitem{EngineeringArray} J. Abraham, et al., 
Nucl. Instrum. Meth. A 523 (2004) 50–95, doi:10.1016/j.nima.2003.12.012.

\bibitem{AugerPrime} A. Castellina for the Pierre Auger Collaboration, 
EPJ Web of Conferences 210, 06002 (2019)

\bibitem{Scalers} The Pierre Auger Collaboration, 
Journ. Instr. 6 (2011) P01003; 
``Scaler rates from the Pierre Auger Observatory: a new proxy of solar activity'', Accepted by Astroph. Journal (2025).

\bibitem{Elves} The Pierre Auger Collaboration, 
Eos 101 Sci. News AGU (2020); 
Earth Space Sci. 7 (2020) e2019EA000582 

\bibitem{TGFs} The Pierre Auger Collaboration, 
PoS(ICRC2021)395;
R. Colalillo for the Pierre Auger Collaboration, 
14 CGEO2: PoS(ICRC2023)439

\bibitem{Aerosols} The Pierre Auger Collaboration, G. Curci, 
Atmos. Res. 149 (2014) 120-135

\bibitem{FRAM} The Pierre Auger Collaboration et. al., 
JINST 16 (2021) P06027; M. Jelínek et. al., 
Astronomy \& Astrophysics 454, 3, (2006) L119 - L122, https://doi.org/10.1051/0004-6361:20065092

\bibitem{AERA} The Pierre Auger Collaboration, 
JINST 7 (2012) P11023; 
JINST 7 (2012) P10011; 
Phys. Rev. D 93 (2016) 122005

\bibitem{GHZ} K. Louedec for the P. Auger Collaboration, 
arXiv:1310.4603 [astro-ph.IM], PoS(EPS-HEP 2013)027

\bibitem{FAST} T. Fujii et. al., 
EPJ Web of Conferences 283, 06010 (2023), https://doi.org/10.1051/epjconf/202328306010

\bibitem{GRAND} https://grand.cnrs.fr/

\bibitem{IceCube} https://icecube.wisc.edu/

\bibitem{LAGO} http://lagoproject.net/ ; I. Sidelnik and H. Asorey for the LAGO Collaboration, 
Nucl. Instr. Meth. A 876 (2017) 173-175,
https://doi.org/10.1016/j.nima.2017.02.069


\bibitem{seismic} S. Spagnotto et. al., 
Journal of South American Earth Sciences 63 (2015) 36-47

\bibitem{Satellites} The Pierre Auger Collaboration et. al., 
Optica 11 (2024) 263-272

\bibitem{DS3} https://www.esa.int/Enabling\_Support/Operations/ESA\_Ground\_Stations/Malarguee\_-\_DSA\_3

\bibitem{OpenData} The Pierre Auger Collaboration, Eur. Phys. J. C 85 (2025) 70

\bibitem{Outreach} F. Convenga for the Pierre Auger Collaboration, ``Education and Outreach Activities within the Pierre
Auger Collaboration'', ICRC 2025 \#797
}
\end{thebibliography}
\end{document}